\documentclass[aps,pra,a4paper,floatfix,showpacs,notitlepage,twocolumn]{revtex4-1} 
\usepackage{amsmath}
\usepackage{amsfonts}
\usepackage{amssymb}
\usepackage{graphicx}
\graphicspath{{./}}
\usepackage{epsfig}
\usepackage{epstopdf}
\usepackage{xcolor}
\usepackage{empheq}
\usepackage{cancel}
\usepackage{braket}

\DeclareMathOperator\arccosh{arccosh}

\newcommand\hp[1]{\hphantom{#1}}

\begin{document}

\title{Mechanical entanglement detection in an optomechanical system}
\author{Francesco Massel}
\email[]{francesco.p.massel@jyu.fi}
\affiliation{Department of Physics and Nanoscience Center, University of Jyvaskyla,
P.O. Box 35 (YFL), FI-40014 University of Jyvaskyla, Finland}

 \begin{abstract}
   We propose here a setup to generate and evaluate the entanglement between two
   mechanical resonators in a cavity optomechanical setting. As in previous
   proposals, our scheme includes two driving pumps allowing for the generation
   of two-mode mechanical squeezing. In addition, we include here four
   additional probing tones, which allow for the separate evaluation of the
   collective mechanical quadratures required to estimate the Duan quantity,
   thus allowing us to infer whether the mechanical resonators are entangled.
\end{abstract}
\maketitle

Since the early years of quantum mechanics, it was realised that some of the
consequences borne from its fundamental principles are in stark contradiction
with an intuitive picture of reality deriving from our daily experience of the
world. Arguably, one of its most unsettling aspects--yet potentially useful in the
manipulation of information at the quantum level--  is represented by
entanglement, which is a form of correlation inherent to quantum
mechanical systems. One of the prototypical examples of entangled systems was
discussed as early as 1935 by Einstein Podolsky and Rosen
\cite{Einstein:1935hx}, in the attempt to prove the incompleteness of quantum
mechanics. In their paper the authors discuss a \textit{gedankenexperiment} in
which they show how the measurement of position or momentum on a quantum system
can affect the state of a second system causally disconnected from the first
one. While in the decades following the paper by Einstein Podolsky and Rosen,
the reality of entanglement has been demonstrated in the context of quantum
optics (see e.g. \cite{Aspect:1982br,Reid:1986bv,Pan:2003kv}), the realisation of the
experiment involving the position and momentum of a (macroscopic) material
system, much along the lines of the original proposal discussed in
\cite{Einstein:1935hx}, has not been carried out.

In the recent years the progress in the physics of optomechanical systems
\cite{Milburn:2012cu, Aspelmeyer:2014ce} has opened the prospect of entangling
the mechanical degrees of freedom of two macroscopic mechanical oscillators
\cite{Mancini:2002cn,Pinard:2005ex,Pirandola:2006hy,Schmidt:2012hsa,Abdi:2012hm,Tan:2013ki,Wang:2013hk,Woolley:2014he,Szorkovszky:2014dr},
following the experimental realisation entanglement between a mechanical
oscillator with a microwave field \cite{Palomaki:2013gs} and the
backaction-evading measurement of collective mechanical modes
\cite{OckeloenKorppi:2016cp}. In particular in
\cite{Wang:2013hk,Woolley:2014he,Li:2015ed} it has been proposed to use a common
optical cavity with an appropriate optical drive in order to entangle the
degrees of freedom of two mechanical resonators. While the realisation of such a
setup is within reach of the current experimental capabilities
\cite{Massel:2012dw,OckeloenKorppi:2016cp}, a reliable characterisation of the
entanglement properties between mechanical resonators, considering the
limitations imposed by the current experimental parameters range, is lacking.
In particular, in the theoretical proposals based on a one-cavity/two-mechanical
resonators concept, the most severe limitation is represented by the
impossibility, in the current experimental settings, of separately addressing
the two mechanical modes. The separate addressability of the mechanical modes,
which, in principle, could be realised by engineering mechanical resonators of
sufficiently different frequency, would open up the prospective of state
tomography for the two mechanical resonators \cite{Vanner:2015im} and thus
provide a different route to the characterisation of the entanglement properties
of the two mechanical resonators. 

\begin{figure}[ht]
\centering
\includegraphics[width=0.9\columnwidth]{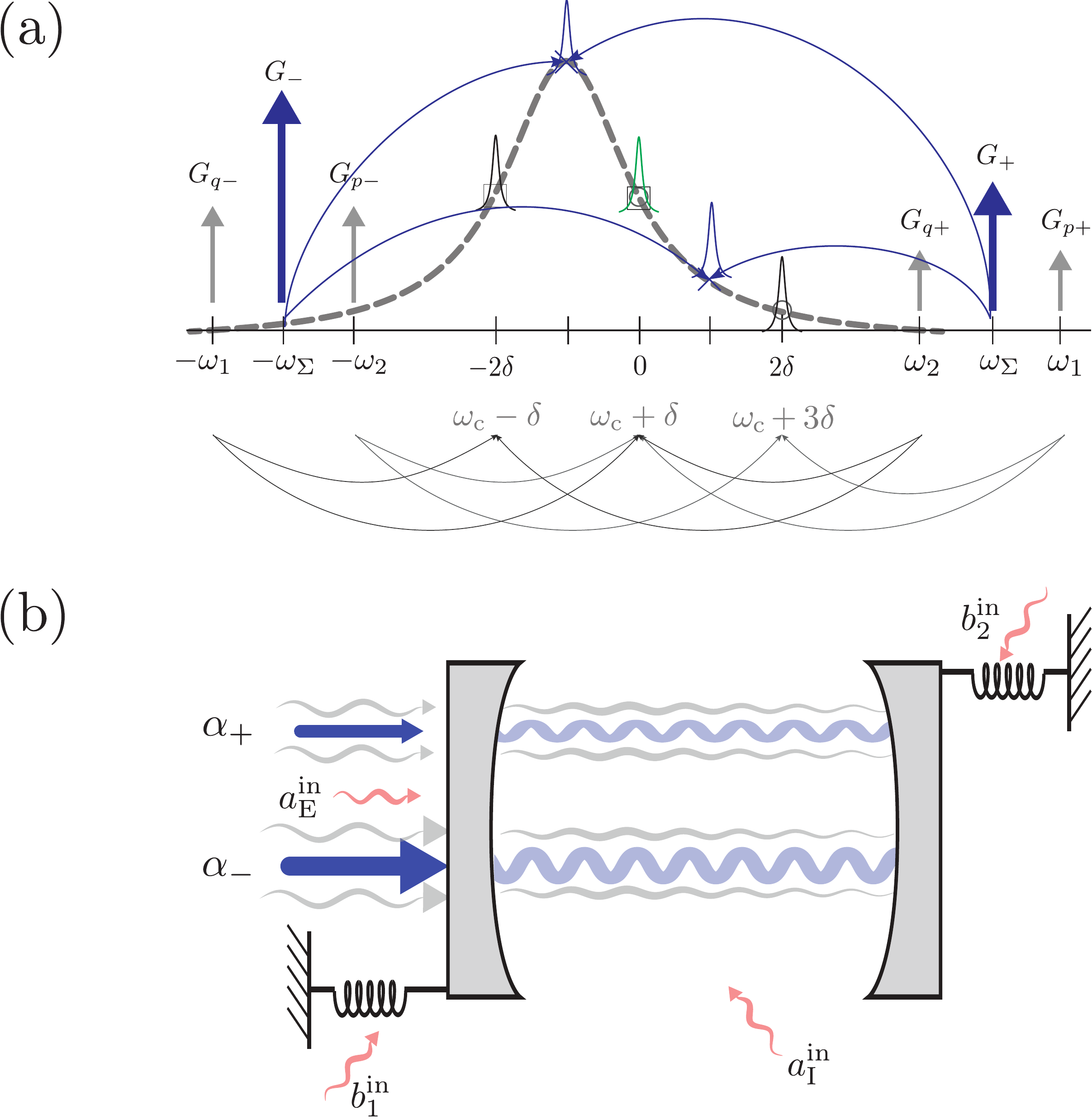}
\caption{\textit{System setup}. (a) Sidebands generated by the pump setup
  discussed in the article (pictorial view) (b) Pumping scheme. The two pumps
  (blue), generate sidebands at $\pm \delta$ ($\omega_\mathrm{c}$ and
  $\omega_\mathrm{c}+2\delta$ in the original frame), while the two probes
  (grey) generate sidebands respectively at $-2\delta$ and $0$
  ($\omega_\mathrm{c}-\delta$ and $0$ in the original frame), and at $0$ and
  $2\delta$ ($\omega_\mathrm{c}+\delta$ and $\omega_\mathrm{c}+3\delta$ in the
  original frame). In this work we will focus on the peak generated at $0$(
  panel (a), green peak).}
\label{fig:freqsch}
\end{figure}

In this article, we propose a detection scheme for currently available
one-cavity/two-resonators setups related to the concept of
backaction-evading measurement
\cite{Caves:1980jpa,Braginsky:1980hj,Clerk:2008je,Clerk:2010dh,Hertzberg:2010ge,Suh:2014gb,OckeloenKorppi:2016cp},
in a system constituted by one resonant cavity coupled to two mechanical
resonators by radiation pressure force (see Fig.~\ref{fig:freqsch}). The scheme
proposed here, based on a novel pump/probe setup, allows us to verify the
entanglement between the two mechanical resonators.

As previously noted in the literature \cite{Woolley:2014he}, it is possible to
induce two-mode squeezing \cite{Walls:2008em} on the mechanical degrees of
freedom by suitably driving the cavity with two coherent tones. While this
two-mode squeezing represents the key element in the definition of the
entanglement between the two mechanical operators, the direct observation of
entanglement between the mechanical resonators has proven elusive. Here we
discuss how, in order to gain access to the different quadratures of the
collective modes needed to determine the entanglement between mechanical
resonators, 4 extra probes are required.  More specifically, the setup proposed
here allows us to infer the value of the Duan quantity \cite{Duan:2000fw} for
the collective quadratures associated with the dynamics of the two mechanical
resonators from the noise spectrum of the output field.

We evaluate here the Duan quantity and the ensuing Duan bound
\cite{Duan:2000fw}, which represent a possible inseparability criterion for a
bipartite system and have been previously considered in the context of
entanglement between mechanica resonators in cavity optomechanics
\cite{Pinard:2005ex,Woolley:2014he,Li:2015ed}.

The Duan bound is expressed in terms of an upper bound to the total variance
of EPR-like observables.  For instance, for two mechanical resonators, the Duan
criterion could be expressed as
$$
 \braket{\Delta X_\Sigma^2} + \braket{\Delta Y_\Delta^2} \leq 1 
$$ 
where $X_\Sigma = X_1 + X_2$ and $Y_\Delta =Y_1- Y_2$, with $X_{1,2}$ and
$Y_{1,2}$ representing two orthogonal quadrature operators( e.g. the position
and momentum operators) associated with the dynamics of subsystem $1$ and $2$
respectively.

\paragraph{Setup and equations of motion--}The setup considered here consists
of two mechanical resonators dispersively coupled to a single optical
cavity. The general Hamiltonian of the (isolated) system can be written as
\begin{align}
  H= \omega_\mathrm{c} a^\dagger a + &\omega_1 b_1^\dagger b_1 + \omega_2 b_2^\dagger b_2 \nonumber \\
       + & g_0 a^\dagger a  \left[ \left(b_1^\dagger+ b_1 \right) + \left(b_2^\dagger +b_2 \right) \right]
 \label{eq:1p}
\end{align}
where $a$, $b_1$ and $b_2$, along with their hermitian conjugates, represent the
field operators associated with the cavity and the two mechanical resonators,
with resonant frequencies $\omega_\mathrm{c}$, $\omega_1$ and $\omega_2$
respectively. Furthermore, we have assumed that each mechanical resonator is
coupled through a radiation-pressure coupling term of strength $g_0$.  In our
scheme the system is strongly driven at frequencies $\omega_\mathrm{c}+\omega_1$
with a coherent pumping tone of amplitude $\alpha_+$ and
$\omega_\mathrm{c}-\omega_2$ with amplitude $\alpha_-$ ($G_\pm=g_0 \alpha_\pm$).
In addition to the driving at $\omega_\mathrm{c}\pm \omega_{1,2}$, the cavity
driving scheme proposed here comprises four extra detection tones at frequencies
$\omega_\mathrm{c}+\omega_1 \pm \delta$ and
$\omega_\mathrm{c}-\omega_2 \pm \delta$, with
$\delta=\left(\omega_1-\omega_2\right)/2$ (see Fig.~\ref{fig:freqsch}).

In the appropriate frame ($\omega_{\mathrm{c}}+\delta$ and
$\omega_{\Sigma}=(\omega_1+\omega_2)/2$ for cavity and mechanics respectively),
after linearisation around the pumping tones, and neglecting terms rotating at
frequencies $\omega \simeq 2 \omega_\Sigma$ since we assume
$\omega_\Sigma \gg \kappa$ (rotating-wave approximation), the Hamiltonian can be
written as
  \begin{align}
    H= &- \delta a^\dagger a + 2 \delta \left(b^\dagger_1 b_1 - b^\dagger_2 b_2 \right) \nonumber \\ 
         + &\left[ G_+ a^\dagger  \left(b^\dagger_1 +b^\dagger_2\right) + G_- a^\dagger \left(b_1 +b_2\right) + \text{h.c.} \right.\nonumber \\
         + & G_{p+} e^{i \delta t} a^\dagger  \left(b^\dagger_1 +b^\dagger_2\right)   + G_{p-} e^{i \delta t} a^\dagger
             \left(b_1 +b_2 \right)+ \text{h.c.} \nonumber \\
          + &\left. G_{q+} e^{-i \delta t} a^\dagger  \left(b^\dagger_1 +b^\dagger_2\right)+ G_{q-} e^{-i \delta t}
            a^\dagger \left(b_1 +b_2\right) + \text{h.c.} \right]
    \label{eq:2p}
  \end{align}  
  with $G_{p,q\pm}= g_0 \alpha_{p,q\pm}$.

  Choosing $G_+/G_-=G_{p+}/G_{p-}=G_{q+}/G_{q-}$, we can write the quantum
  Langevin equations (QLEs) associated with the linearized system Hamiltonian
  \eqref{eq:2p}, in the Fourier domain, as (see \ref{sec:EOMS})
\begin{align}
 \chi^{-1}_{\mathrm{c}}\left(\omega+\delta\right) a_{\omega \phantom{,1}} =&
 - i \mathcal{G} \left[\beta_{1\,\omega} +\beta_{2\,\omega} \right] \nonumber \\ 
 &- i \mathcal{G}_{\Delta1}  \left[\beta_{1\,\omega-\delta} +\beta_{2\,\omega-\delta} \right] \nonumber \\  
 & - i \mathcal{G}_{\Delta2} \left[\beta_{1\,\omega+\delta} +\beta_{2\,\omega+\delta} \right] 
+ \sqrt{\kappa} a^{\mathrm{in}}_\omega \nonumber \\
\nonumber \\
\chi^{-1}_{\mathrm{m}}\left(\omega-\delta \right) \beta_{\omega,1} = &
 - i \mathcal{G}^* a_\omega 
 - i \mathcal{G}^*_{\Delta1}  a_{\omega-\delta} \nonumber \\
 &- i \mathcal{G}^*_{\Delta2}  a_{\omega+\delta} + \sqrt{\gamma} \beta^{\mathrm{in}}_{1\,\omega}\nonumber \\
\nonumber \\
\chi^{-1}_{\mathrm{m}}\left(\omega+\delta \right) \beta_{2\,\omega} = & 
 - i \mathcal{G}^* a_\omega 
 - i \mathcal{G}^*_{\Delta1}  a_{\omega-\delta} \nonumber \\
 &- i \mathcal{G}^*_{\Delta2}  a_{\omega+\delta} + \sqrt{\gamma} \beta^{\mathrm{in}}_{2\,\omega} 
 \label{eq:3p}
\end{align}
where
$\mathcal{G}=\sqrt{\left|G_-\right|^2-\left|G_+\right|^2}$, 
$\mathcal{G}_{\Delta1}=\left[G_- G_{p-}^*-G_+ G_{p+}^*\right]/\mathcal{G}$, 
$\mathcal{G}_{\Delta2}=\left[G_- G_{q-}^*-G_+ G_{q+}^*\right]/\mathcal{G}$ and
$\chi_\mathrm{x}(\omega)^{-1}=\gamma_\mathrm{x}/2-i \omega$ ($\mathrm{x}=1,2$,
$\gamma_1=\gamma$,$\gamma_2=\kappa$).
The input field introduced in \eqref{eq:3p} include contributions both from
internal and external noise (see Fig.~\ref{fig:freqsch})
$$
 \sqrt{\kappa} a^{\mathrm{in}}_\omega= \sqrt{\kappa_\mathrm{i}}
 a^{\mathrm{in}}_{\mathrm{i}\,\omega} +
\sqrt{\kappa_\mathrm{e}} a^{\mathrm{in}}_{\mathrm{e}\,\omega}.
$$
Most importantly, the mechanical operators 
\begin{align}
  \beta_{1\,\omega} = u b_{1\,\omega} +v b_{2\,-\omega}^\dagger, \nonumber \\ 
  \beta_{2\,\omega} = u b_{2\,\omega} +v b_{1\,-\omega}^\dagger,
  \label{eq:12}
\end{align}
where $u=G_+/\sqrt{G_+^2-G_-^2}$, $v=G_-/\sqrt{G_+^2-G_-^2}$, can be regarded as
being generated by the action of the two-mode squeezing operator 
\begin{align*}
  S(r)=\exp\left(  r\, b_{\omega,1} b_{\omega,2} - \mathrm{h.c.}\right)
\end{align*}
with $r=\arccosh u$. $S(r)$ can be shown to give rise to quantum correlations
between the mechanical modes, analogously to the situation encountered in the
context of quantum optics, e.g. in the case of non-degenerate parametric
amplification \cite{Walls:2008em} and thus represents the key ingredient for
entanglement generation in the present setup \cite{Woolley:2014he}.

Our strategy in the solution of the problem in the presence of both pumping and
probing tones, consists now in assuming that
$\left|\mathcal{G} \right|\gg \left|\mathcal{G}_{\Delta1}\right|,
\left|\mathcal{G}_{\Delta2}\right|$, and $\gamma\ll \delta$, treating the
probing tones as a perturbation with respect to the driving tones (see
\ref{sec:equat-moti-pert}). These conditions allow us to solve for the dynamics
of the mechanical resonators as if it was determined by the pumping tone only
and independently for each mode.
\begin{figure}[ht]
\centering
\includegraphics[width=0.9\columnwidth]{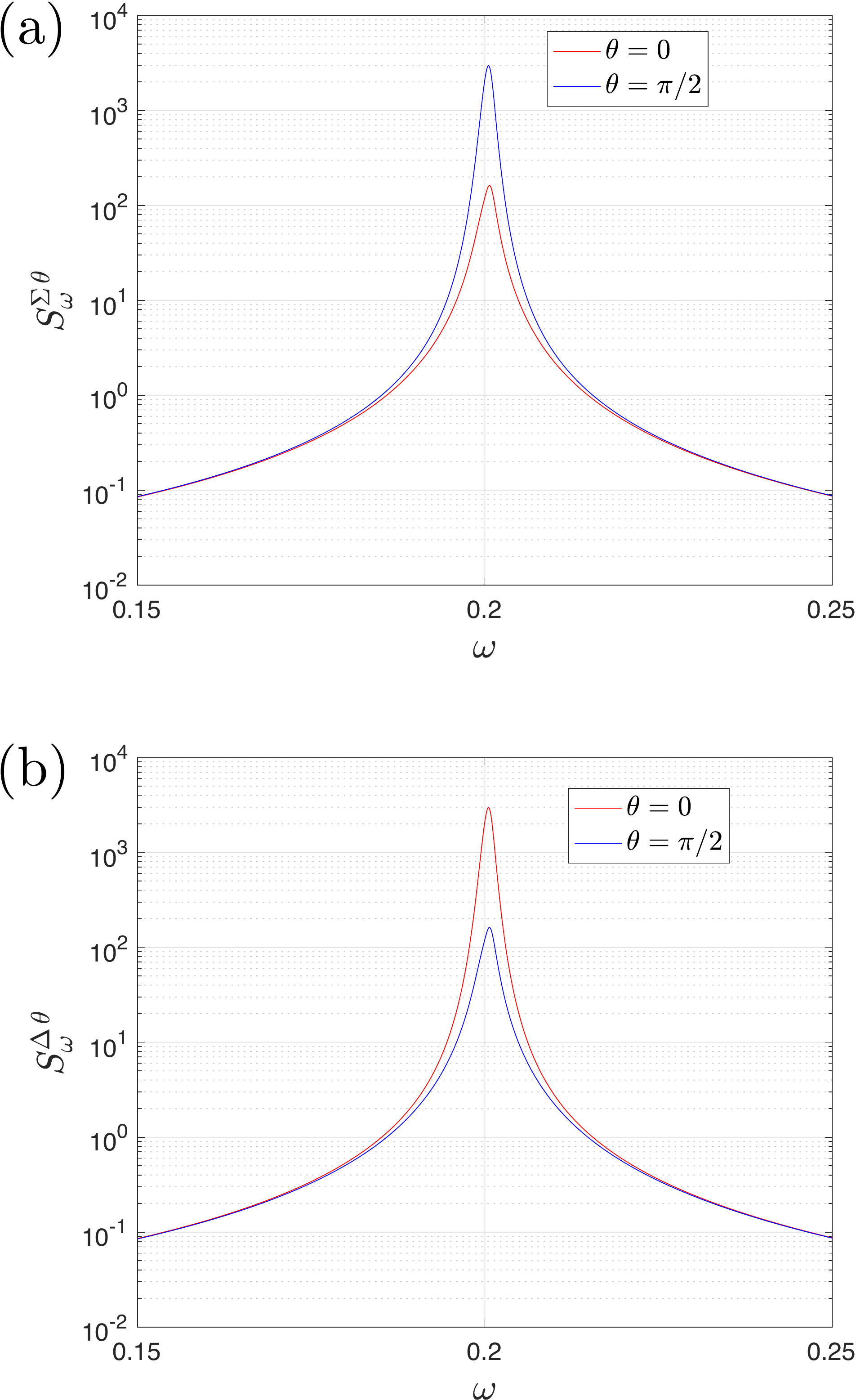}
\caption{\textit{Mechanical spectrum}. (a) Spectrum for the symmetrical
  mechanical quadratures $S_\omega^{\Sigma\,\theta}$, for $\theta=0$, (red) and
  $\theta=\pi/2$ (blue). (b) Spectrum for the antisymmetrical mechanical
  quadratures $S_\omega^{\Delta\,\theta}$ $\theta=0$, (red) and $\theta=\pi/2$
  (blue). System parameters: $\delta=0.2$, $G_-=4.8 \cdot 10^{-2}$,
  $G_+=4.0 \cdot 10^{-2}$, $\gamma=1 \cdot 10^{-5}$, $n_{\mathbf{m}}=10$,
  $n_{\mathbf{c}}^{\mathrm{I}}=n_{\mathbf{c}}^{\mathrm{E}}=0.1$, all energies in
  units of $\kappa$.}
\label{fig:mech_resp}
\end{figure}
In Fig.~\ref{fig:mech_resp}, we have depicted the mechanical noise spectrum,
with the following definitions
\begin{align}
  S_\omega^{\Sigma\,\theta} = \frac{1}{2}\braket{\left\{
                       \braket{X_{-\omega}^{\Sigma\,\theta},X_\omega^{\Sigma\,\theta}}\right\}}
  \nonumber \\
  S_\omega^{\Delta\,\theta} = \frac{1}{2}\braket{\left\{
                       \braket{X_{-\omega}^{\Delta\,\theta},X_\omega^{\Delta\,\theta}}\right\}}
 \label{eq:2}
\end{align}
and
\begin{align}
  X^\Sigma_{\omega} \doteq  X_{1\,\omega} + X_{2\,\omega}\nonumber \\
  X^\Delta_{\omega} \doteq  X_{1\,\omega} - X_{2\,\omega}
  \label{eq:3nn}  
\end{align}

The spectra in Fig.~\ref{fig:mech_resp} are obtained in the presence of thermal noise both
for the cavity ( $\braket{a^\mathrm{in}_{\mathrm{I}\,\omega}
  {a^{in}_{\mathrm{I}\,\omega'}}^\dagger}=\left(n_\mathrm{i}+1\right)\delta_{\omega,\omega'}$,
 $\braket{a^\mathrm{in}_{\mathrm{E}\,\omega}
  {a^{in}_{\mathrm{E}\,\omega'}}^\dagger}=\left(n_\mathrm{e}+1\right)\delta_{\omega,\omega'}$),
and the mechanical bath
($\braket{b^\mathrm{in}_{\mathrm{x}\,\omega}
  {b^{in}_{\mathrm{x}\,\omega'}}^\dagger}=\left(n_\mathrm{m}+1\right)\delta_{\omega,\omega'}$,
$\mathrm{x}=1,2$).

The solution of the equations of motion for the mechanical degrees of freedom
determines the dynamics of the cavity field, giving rise to the appearance five
peaks in the cavity (and output) spectrum. Due to the small value of the
mechanical linewidth, it is possible to consider separately each peak induced in
the cavity field by the mechanical resonators.  In our analysis, we are
interested in particular in the peak at $\omega=0$
($\omega= \omega_\mathrm{c}+\delta$ in the original frame), which comprises
contributions from the dynamics of both mechanical resonators and, as we will
show, contains all the information needed to evaluate the Duan bound. Since for
$\omega \simeq 0$ the mechanical contributions to the cavity field are
predominantly provided by $\mathcal{G}_{\Delta1}\beta_{\omega-\delta,2}$ and
$\mathcal{G}_{\Delta2}\beta_{\omega+\delta,1}$, due to the resonance condition
in the mechanical equations of motion: for $\omega \simeq 0$, $\beta_{\omega,1}$
is resonant at $\omega+\delta$ and $\beta_{\omega,2}$ at $\omega-\delta$.

\paragraph{Output spectrum--}If homodyne detection is performed on the
fluctuations, from Eq.~\eqref{eq:5p}, expressing $\beta_1$ and $\beta_2$ in
terms of $b_1$ and $b_2$, it is possible to monitor the dynamics of the
collective mechanical modes through the measurement of the quadratures of the
output field, namely (for $\omega\simeq 0$, $\delta \ll \kappa$)
\begin{align}
  X_\omega^{\mathrm{out}} &\doteq \left(a^\dagger e^{i \theta}+  a e^{-i \theta}\right)/\sqrt{2}
                            =\nonumber \\ 
                  \sqrt{\kappa} & \left|\chi_\mathrm{c}\left(\omega+\delta\right) \right| 
                     \mathcal{G}_{D} 
                \Big[ 
                    (u+v) \cos\theta 
                      \left(\cos \varphi \bar{X}^\Sigma_{\omega} - \sin \varphi \bar{Y}^\Delta_{\omega} \right) 
                   \nonumber \\
                   + (u-v)& \sin \theta 
                      \left(\cos \varphi \bar{Y}^\Delta_{\omega} + \sin \varphi \bar{X}^\Delta_{\omega} \right)
                   \Big]  + \left[\kappa \chi_\mathrm{c}\left(\omega+\delta\right)-1\right]
                                X^{\mathrm{in}}_{\omega} 
  \label{eq:4p}
\end{align}
where the dynamics of the collective mechanical modes is encoded in the
frequency-shifted quadrature operators $ \bar{X}^\Sigma_{\omega}$, $
\bar{Y}^\Sigma_{\omega}$ and $ \bar{X}^\Delta_{\omega}$, $
\bar{Y}^\Delta_{\omega}$ defined by    
\begin{align}
  \bar{X}^\Sigma_{\omega} \doteq  \bar{X}_{1\,\omega} + \bar{X}_{2\,\omega}\nonumber \\
  \bar{X}^\Delta_{\omega} \doteq  \bar{X}_{1\,\omega} - \bar{X}_{2\,\omega}
    \label{eq:5p}
\end{align}
with
\begin{alignat}{4}
&\bar{X}_{\mathrm{1}\,\omega} &&\doteq \left(b^\dagger_{\mathrm{1}\,-\omega+\delta} +
b_{\mathrm{1}\,\omega+\delta}\right)/\sqrt{2} \nonumber \\
&\bar{X}_{\mathrm{2}\,\omega} &&\doteq  \left(  b^\dagger_{\mathrm{2}\,-\omega-\delta} +
  b_{\mathrm{2}\,\omega-\delta}\right)/\sqrt{2},
\label{eq:5p2}
\end{alignat}
with analogous definitions holding for the quadratures $\bar{Y}_{1\,\omega}$ and
$\bar{Y}_{2\,\omega}$.  Note that the frequency-shifted quadrature operators
defined above, do not directly correspond to the usual quadrature operators. In
the time domain, the operators defined in \eqref{eq:5p2} acquire a nontrivial
time dependence, for instance we have that
 $$
   \bar{X}_{1\,t}= b_\mathrm{t} e^{i \delta t} + b^\dagger_\mathrm{t} e^{-i \delta t}
 $$
and thus correspond to the mechanical quadratures for $\delta=0$ only. While one
cannot directly relate the frequency-shifted quadrature operators to the
regular ones, for each pair of orthogonal quadratures,
the uncertainties associated with the collective quadratures of the mechanical
motion, fulfil the following relation
\begin{align}
  \braket{\Delta \bar{X}_\Sigma^2} + \braket{\Delta \bar{Y}_\Delta^2} =  \braket{\Delta X_\Sigma^2} + \braket{\Delta Y_\Delta^2},  
  \label{eq:6p}
\end{align}
which, crucially, allows us to establish the link between the output spectrum
and the spectrum of the collective mechanical quadratures.
Therefore, the knowledge of one pair of orthogonal frequency-shifted mechanical
quadratures allows one to deduce the value of the corresponding pair of regular
quadratures and, consequently, to infer the value of the Duan quantity from the
spectrum of the output field.

In the derivation of Eq.~\eqref{eq:4p}, we have chosen the
average phase of the detection tones as the phase reference with respect to
which both the phases of the probing tones $\mathcal{G}_{\Delta_1}$ and
$\mathcal{G}_{\Delta_2}$ ($\pm \varphi$) as well as the homodyne detection phase
$\theta$ are referred.  From Eq.~\eqref{eq:4p}, it is possible to relate the
output spectrum $S^{\mathrm{out}\, \theta}_{\omega}$ to the spectrum of the
frequency-shifted mechanical quadratures as
\begin{align}
  &S^{\mathrm{out}\, \theta}_{\omega} =\mathcal{G}^2_D
  \left|\chi_\mathrm{c}(\delta)\right|^2 \\ \nonumber 
    & \left\{
            \left(u+v\right)^2 
          \left[
               \left(\cos \theta \cos \varphi \right)^2  \bar{S}_\omega^{\Sigma\,0} +
               \left(\cos \theta \sin \varphi \right)^2  \bar{S}_\omega^{\Delta\,\pi/2}
          \right] \right. \nonumber \\ 
        & \left.  + \left(u-v\right)^2  
          \left[
                 \left(\sin \theta \cos \varphi \right)^2  \bar{S}_\omega^{\Sigma\,\pi/2} +
               \left(\sin \theta \sin \varphi \right)^2    \bar{S}_\omega^{\Delta\,0}
          \right]
    \right\} \nonumber \\ 
   &+ B^\mathrm{in}+ C_\omega^\mathrm{in} \cos \left[2 \phi \right]
    \label{eq:5p3}    
\end{align}
(see the \ref{sec:duan-quantity-from} for the derivation of the output noise spectrum
$S^{\mathrm{out}\, \theta}_{\omega}$, and the definitions of $B^\mathrm{in}$ and
$C_\omega^{\mathrm{in}}$).
\begin{figure}[ht]
\centering
\includegraphics[width=0.9\columnwidth]{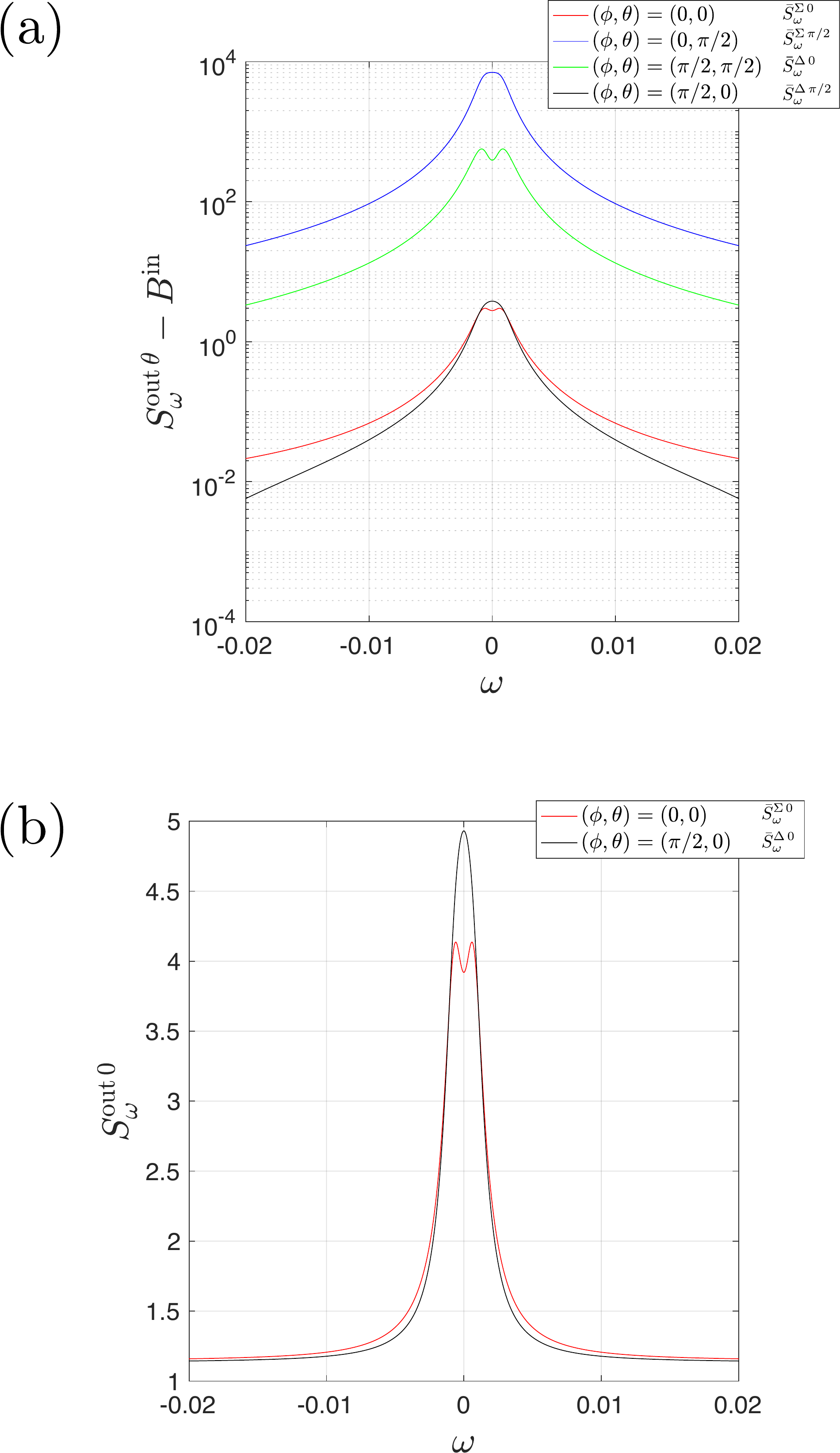}
\caption{\textit{Output spectrum}. (a) Spectrum for the output field for the
  four relevant combinations of $\left(\phi,\theta \right)$. Note the two-mode
  squeezing effect, indicated by the difference in the noise spectra for
  $\left(0,0\right)$ and $\left(0,\pi/2\right)$, and $\left(\pi/2,0\right)$ and
  $\left(\pi/2,\pi/2\right)$, respectively. (b) Same as in (a) focus on the
  spectra for $\left(\phi,\theta \right)=(0,0)$ and
  $\left(\phi,\theta\right)=(\pi/2,0)$ --note here the linear scale;the base
  level corresponds to the pure cavity response. For each value of the pair
  $\left(\phi,\theta\right)$, we have indicated the corresponding shifted
  mechanical noise spectrum. Physical parameters same as in
  Fig. \ref{fig:mech_resp}.}
\label{fig:Outsp}
\end{figure}
Eq.~\eqref{eq:5p} expresses the possibility to access the collective mechanical
noise spectra $\bar{S}_\omega^{\Sigma\,0} $, $\bar{S}_\omega^{\Sigma\,\pi/2} $,
$\bar{S}_\omega^{\Delta\,0} $, $\bar{S}_\omega^{\Delta\,\pi/2} $, by changing
the relative phase of the detection tones $\varphi$ and the phase of the
homodyne detector $\theta$.

The measurement strategy leading to the determination of the Duan quantity
consists in the measurement of the output spectrum for four different
values of $\left(\theta, \varphi\right)=\left(0, 0\right)$,
$\left(0, \pi/2\right)$, $\left(\pi/2,0\right)$, $\left(\pi/2,\pi/2\right)$,
yielding
\begin{align}
     & \left. S^{\mathrm{out}\, 0}_{\omega}\right|_{\varphi=0} &= 
     & \left(u+v\right)   \bar{S}_\omega^{\Sigma\,0} +C_\omega^{\mathrm{in}}  \nonumber \\
     &  \left. S^{\mathrm{out}\, 0}_{\omega}\right|_{\varphi=\pi/2} &= 
     & \left(u+v\right)   \bar{S}_\omega^{\Delta\,\pi/2} -C_\omega^{\mathrm{in}} \nonumber \\
     & \left. S^{\mathrm{out}\, \pi/2}_{\omega}\right|_{\varphi=0} &= 
     & \left(u-v\right)   \bar{S}_\omega^{\Sigma\,\pi/2} +C_\omega^{\mathrm{in}}\nonumber \\
     & \left. S^{\mathrm{out}\, \pi/2}_{\omega}\right|_{\varphi=\pi/2}& = 
     & \left(u-v\right)   \bar{S}_\omega^{\Delta\,0} -C_\omega^{\mathrm{in}}.
      \label{eq:1nn}
\end{align}
Having in mind the relation established by Eq.~\eqref{eq:6p} The Duan quantity
can then be determined as the sum of the appropriate output quadrature spectra
as determined in Eqs. \eqref{eq:1nn}. The output spectrum therefore provides a
measurement of the collective mechanical quadratures induced by the pumping
tones $\alpha_\pm$, disregarding higher-order effects induced by the detection
tones. As detailed in \ref{sec:equat-moti-pert}, for $\omega \simeq 0$ the
correction to the dynamics of the mechanical resonators due to the detection
tones is of the order $\left(\mathcal{G}_D/\mathcal{G}\right)^2$, and thus can
be safely neglected for a suitable choice of readout tones.

\begin{figure}[ht]
\centering
\includegraphics[width=0.9\columnwidth]{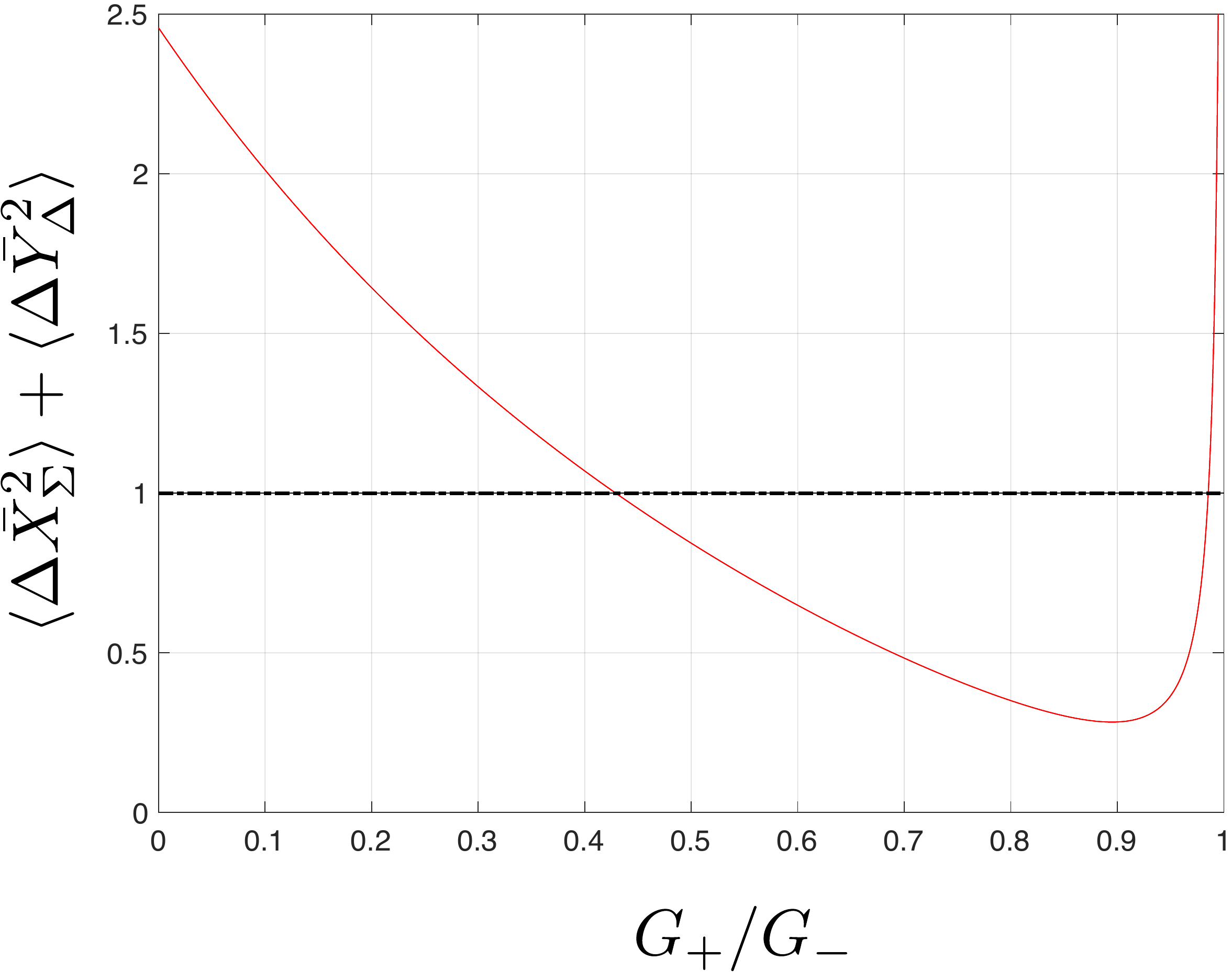}
\caption{Duan quantity as obtained from the shifted mechanical operators as a
  function of the ratio $G_+/G_-$, (with $G_-$ kept fixed), showing that for a
  ratio $G_+/G_-$ between $\simeq 0.45$ and $\simeq 0.99$, the mechanical
  resonators are entangled . Other physical parameters same as in
  Fig. \ref{fig:mech_resp}}.
\label{fig:Duan_q}
\end{figure}
In the limit $\kappa \gg \delta, \mathcal{G} \gg \gamma$, the quantity given in
Eq.~\eqref{eq:6p} can be written as
\begin{align}
  & \braket{\Delta \bar{X}_\Sigma^2} + 
  \braket{\Delta \bar{Y}_\Delta^2} =\nonumber \\
  &\left[ \gamma \kappa\left( n_\mathrm{m}+\frac{1}{2}\right)+
          4\left(G_-^2+G_+^2\right) \left( n_\mathrm{c}+\frac{1}{2}\right)
  \right]/\left[2 (G_-^2-G_+^2)\right]
  \label{eq:7p}
\end{align}
(see \ref{sec:duan-quantity-from} for the full exact expression), which corresponds
to the approximate formula given in \cite{Woolley:2014he} for the Duan quantity
in the adiabatic limit.  

The detection scheme proposed here represents a somewhat idealised setup. A
first apporximation in our approach is represented by the rotating-wave
approximation. Considering that that discarded terms in performing the rotating
wave approximation are of the order on
$\sim \left(\kappa/\omega_\mathrm{m}\right)^2$, reaching a range of experimental
parameters in which this approximation does not seem to represent an
insurmountable experimental challenge (see e.g. \cite{Pirkkalainen:2015ki},
where $\kappa/\omega_\mathrm{m} \simeq 0.05$).

The most relevant source of non-ideality is probably represented by the presence
of parametric modulation terms to the dynamics of the cavity. This effect was
observed in the context of the optomechanical generation and detection of
mechanical squeezing \cite{Wollman:2015gx,Pirkkalainen:2015ki} and is likely to
affect the detection of mechanical entanglement as well.
 
Among other non-idealities likely encountered in the specific experimental
realisation of the detection scheme discussed here, it is worth mentioning that
the two mechanical resonators are inevitably not identical and the thermal baths
to which they are coupled are not necessarily at the same temperature. The most
severe constraint posed by the difference between the two mechanical resonators
seems to be posed by the different value of the bare optomechanical coupling
$g_0$, which however can be compensated by a suitably engineered pumping scheme
(see e.g. \cite{Woolley:2014he}).

We have proposed here a potential scheme for the detection of entanglement
between mechanical resonators coupled to a common optical cavity, establishing a
straightforward setup for the quantification of such entanglement in terms of
the Duan quantity.  Its feasibility is within the framework of current
experimental capabilities.

The author would like to thank Mika Sillanp\"a\"a and Tero Heikkil\"a for useful
discussions. This work was supported by the Academy of Finland (Contract
No. 27545).
  
\appendix

\section{Derivation of the equations of motion}
\label{sec:EOMS}

In the appropriate frame ($\omega_{\mathrm{c}}+\delta$ and $\omega_{\Sigma}=(\omega_1+\omega_2)/2$ for cavity and
  mechanics respectively), and invoking the rotating-wave approximation, the Hamiltonian given in Eq.~(2) of the main
  text can be written as
  \begin{align}
    H= &- \delta a^\dagger a + \delta \left(b^\dagger_1 b_1 - b^\dagger_2 b_2 \right) \nonumber \\ 
         + &\left[ G_+ a^\dagger  \left(b^\dagger_1 +b^\dagger_2\right) + G_- a^\dagger \left(b_1 +b_2\right) + \text{h.c.} \right.\nonumber \\
         + & G_{p+} e^{i \delta t} a^\dagger  \left(b^\dagger_1 +b^\dagger_2\right)   + G_{p-} e^{i \delta t} a^\dagger
             \left(b_1 +b_2 \right)+ \text{h.c.} \nonumber \\
          + &\left. G_{q+} e^{-i \delta t} a^\dagger  \left(b^\dagger_1 +b^\dagger_2\right)+ G_{q-} e^{-i \delta t}
            a^\dagger \left(b_1 +b_2\right) + \text{h.c.} \right]
    \label{eq:2s}
  \end{align}
with $\delta=\left(\omega_1-\omega_2\right)/2$. 
From the Hamiltonian \eqref{eq:2s}, it is possible to write down the quantum Langevin equations for cavity and mechanical degrees of
freedom as
\begin{align}
  \dot{a} = &-i  \delta a  -i \sqrt{2}
                         \left[\left( G_+ b^\dagger_\Sigma + G_- b_\Sigma \right) \right.\nonumber  \\
 &\hp{-i  \delta a  -i \sqrt{2}}
                             +  \left( G_{p+} e^{i \delta t}  b_\Sigma^\dagger+  G_{p-} e^{i \delta t} b_\Sigma \right)\nonumber  \\
 &\hp{-i  \delta a  -i \sqrt{2}}   
                      \left.+\left( G_{q+} e^{i \delta t} b_\Sigma^\dagger + G_{q-} e^{i \delta t} b_\Sigma
                              \right)\right] \nonumber  \\
                        \nonumber  \\
   \dot{b}_\Sigma = & -i  \delta b_\Delta -i\sqrt{2}
                                      \left[ \left( G_+ a^\dagger+ G^*_- a \right) \right.\nonumber \\
&\hp{-i  \delta  b_\Delta  -i \sqrt{2}}
                              + \left( G_{p+} e^{i \delta t} a^\dagger+ G^*_{p-} e^{-i \delta t} a \right)\nonumber  \\
&\hp{-i  \delta  b_\Delta  -i \sqrt{2}}
                       \left.+ \left( G_{q+} e^{-i \delta t} a^\dagger+ G^*_{q-} e^{i \delta t} a  \right)\right]
                                \nonumber  \\
                        \nonumber  \\
   \dot{b}_\Delta= & -i  \delta b_\Sigma 
  \label{eq:3s}  
\end{align}
where we have defined $b_\Sigma =\left( b_1 +b_2 \right)/\sqrt{2}$, $b_\Delta =\left( b_1 - b_2 \right)/\sqrt{2}$. 
If  we now  write the EOM for the linear
combination $\left(G_- b_{\Sigma,\Delta} + G_+ b_{\Sigma,\Delta}^\dagger\right)$, from Eq.~\eqref{eq:3s} we can write
\begin{align}
   &\left(G_- \dot{b}_\Sigma + G_+ \dot{b}_\Sigma^\dagger\right) =\nonumber \\
    &-i  \delta \left( G_- b_\Delta - G_+ b_\Delta^\dagger  \right) \nonumber \\
   &-i \sqrt{2} \left[ \left( \left| G_- \right|^2 a  - \left| G_+ \right|^2 a + 
         \cancel{G_- G_+ a^\dagger} - \cancel{G_- G_+ a^\dagger} \right)   \right.   \nonumber \\ 
&\hp{-i \sqrt{2} \Big[}
    +\left(  G_- G_{p-}^* e^{-i \delta t} a   -  G_+ G_{p+} e^{-i\delta t}a \right. \nonumber \\
  &\hp{-i \sqrt{2} \Big[\Big[\Big[} \left.       
  + \cancel{G_- G_{p+} e^{i\delta t} a^\dagger} - \cancel{G_+ G_{p-}e^{i\delta t}a^\dagger}\right) 
\nonumber \\ 
 &\hp{-i \sqrt{2} \Big[} 
                + \left(  G_- G_{q-}^* e^{i \delta t} a -  G_+ G_{q+} e^{i\delta t}a \right. \nonumber \\
 & \hp{-i \sqrt{2} \Big[\Big[\Big[} \left. \left. 
               + \cancel{G_- G_{q+} e^{-i\delta t} a^\dagger} - \cancel{G_+ G_{q-} e^{-i\delta t}a^\dagger}  \right)  \right].
 \label{eq:4s}
\end{align}
and, analogously,
\begin{align}
  \left(G_- \dot{b}_\Delta - G_+ \dot{b}_\Delta^\dagger\right) = -i \delta \left( G_- b_\Sigma + G_+ b_\Sigma^\dagger \right)
  \label{eq:5s}
\end{align}
The cancellations on the second and third line of Eq.~\eqref{eq:4s}, needed to
recast the problem in terms of Bogolyubov modes, occur only if
$G_{p+}/G_{p-}=G_{q+}/G_{q\,-}=G_+/G_-$. In this case,
\begin{align}
  \dot{a} =&-i \delta a\nonumber \\
               & -i \sqrt{2} \left[ \mathcal{G}^* + \mathcal{G}^*_{\Delta1} e^{i \delta t}+\mathcal{G}^*_{\Delta2}
                         e^{-i \delta t}\right]  \beta_\Sigma \nonumber \\
  \dot{\beta}_\Sigma =&-i \delta \beta_\Delta \nonumber \\ 
                &-i\sqrt{2} \left[ \mathcal{G}  +\mathcal{G}_{\Delta1} e^{-i \delta t}  +
  \mathcal{G}_{\Delta2}e^{i \delta t} \right] a \nonumber \\
  \dot{\beta}_\Delta =& -i \delta \beta_\Sigma
\label{eq:6s}
\end{align}
where we have adopted the definitions 
\begin{align}
  &\beta_\Sigma = u b_\Sigma + v {b_\Sigma}^\dagger \nonumber \\
  &\beta_\Delta  =u b_\Delta - v {b_\Delta}^\dagger \nonumber \\
  &\mathcal{G}=\sqrt{\left|G_-\right|^2-\left|G_+\right|^2} \nonumber \\
  &\mathcal{G}_{\Delta1}=\left[G_- G_{p-}^*-G_+ G_{p+}^*\right]/\mathcal{G} \nonumber \\
  &\mathcal{G}_{\Delta2}=\left[G_- G_{q-}^*-G_+ G_{q+}^*\right]/\mathcal{G}. 
 \label{eq:7s}
\end{align}
In order to ensure that the mechanical operators on the rhs of Eq.~\eqref{eq:3s}
can be all expressed in terms of the same Bogolyubov operator, we have to verify
under what conditions the following three (in principle different)
transformations
\begin{align}
 & G_- b_\Sigma + G_+ b^\dagger_\Sigma = \mathcal{G} \left( u b_\Sigma + v b^\dagger_\Sigma \right) \nonumber \\
 & G_{p-} b_\Sigma + G_{p+} b^\dagger_\Sigma = \mathcal{G}_{\Delta1} \left( u' b_\Sigma + v' b^\dagger_\Sigma \right)
  \nonumber \\
 &G_{q-} b_\Sigma + G_{q+} b^\dagger_\Sigma = \mathcal{G}_{\Delta2} \left( u'' b_\Sigma + v'' b^\dagger_\Sigma \right) \nonumber \\
  \label{eq:8s}
\end{align}
coincide, having set
\begin{alignat}{4}
    &u\hp{''}=\frac{G_-}{\mathcal{G}},                   &\quad&  v\hp{''}=\frac{G_+}{\mathcal{G}}  \nonumber\\
    &u'\hp{'}=\frac{G_{p-}}{\mathcal{G}_{\Delta1}},   &\quad&  v'\hp{'}=\frac{G_{p+}}{\mathcal{G}_{\Delta1}}  \nonumber\\
    &u''=\frac{G_{q-}}{\mathcal{G}_{\Delta2}},  &\quad&  v''=\frac{G_{q+}}{\mathcal{G}_{\Delta2}}. 
\label{eq:9s}
\end{alignat}
It turns out that the condition
$G_{p+}/G_{p-}=G_{q+}/G_{q\,-}=G_+/G_-$ is a sufficient condition to determine $u=u'=u''$ and $v=v'=v''$.
In our analysis we have set the reference phase to be given by the pump tones $G_+$ and $G_-$. 
Focusing on $u'$, we have
\begin{align}
  \frac{\mathcal{G}^2}{\left|G_+\right|}&=\left|G_+ - G_- \frac{G_-^*}{G_+^*}\right| =\left|G_+ - G_- \frac{G_{p-}^*}{G_{p+}^*}\right| \nonumber \\
                                                       &=\left|\frac{G_+ G^*_{p+}-G_- G^*_{p-}}{G_{p+}^*}\right| \nonumber \\
                                                       &=\left|\frac{\mathcal{G}_{\Delta1}}{G_{p+}^*}\right| \implies u=u' 
  \label{eq:10s}
\end{align}
and analogously for $v'$, $u''$ and $v''$.
Defining
\begin{align}
 \beta_{1} = & \frac{\beta_\Sigma + \beta_\Delta}{\sqrt{2}} =u b_{1} +v b_{2}^\dagger \nonumber \\ 
 \beta_{2} = & \frac{\beta_\Sigma - \beta_\Delta}{\sqrt{2}} = u b_{2} +v b_{1}^\dagger,
   \label{eq:11s}
\end{align}
we can write the quantum Langevin equations of motion induced by the
Hamiltonian \eqref{eq:2s} in the Fourier domain as
 \begin{align}
   & {\left(\chi^{\mathrm{c}}_{\omega+\delta}\right)}^{-1} a_\omega =
  - i \mathcal{G} \left[\beta_{1\,\omega} +\beta_{2\,\omega} \right]  \nonumber    \\
&\hp{{(\chi^{\mathrm{c}}_{\omega+\delta})}^{-1} a_\omega =}
  - i \mathcal{G}_{\Delta1}  \left[\beta_{1\,\omega-\delta} +\beta_{2\,\omega-\delta} \right] \nonumber    \\
&\hp{{(\chi^{\mathrm{c}}_{\omega+\delta})}^{-1} a_\omega =}
  - i \mathcal{G}_{\Delta2} \left[\beta_{2\,\omega+\delta} +\beta_{2\,\omega+\delta} \right] \nonumber \\
  &\hp{{(\chi^{\mathrm{c}}_{\omega+\delta})}^{-1} a_\omega =}
 + \sqrt{\kappa_{\mathrm{e}}} a^{\mathrm{in}}_{\mathrm{E}\,\omega} +  \sqrt{\kappa_{\mathrm{i}}} a^{\mathrm{in}}_{\mathrm{I}\,\omega}\nonumber \\
 \nonumber \\
&{\left(\chi^{\mathrm{m}}_{\omega-\delta}\right)}^{-1} \beta_{1\,\omega} = 
 - i \mathcal{G}^* a_\omega \nonumber    \\
&\hp{{\left(\chi^{\mathrm{m}}_{\omega+\delta}\right)}^{-1} \beta_{2\,\omega}=}
 - i \mathcal{G}^*_{\Delta1}  a_{\omega-\delta}
 - i \mathcal{G}^*_{\Delta2}  a_{\omega+\delta} + \sqrt{\gamma} \beta^{\mathrm{in}}_{1\,\omega}\nonumber \\
 \nonumber \\
 &{\left(\chi^{\mathrm{m}}_{\omega+\delta}\right)}^{-1} \beta_{2\,\omega} = 
  - i \mathcal{G}^* a_\omega \nonumber    \\
&\hp{{\left(\chi^{\mathrm{m}}_{\omega+\delta}\right)}^{-1} \beta_{2\,\omega}=}
  - i \mathcal{G}^*_{\Delta1}  a_{\omega-\delta}
  - i \mathcal{G}^*_{\Delta2}  a_{\omega+\delta} + \sqrt{\gamma} \beta^{\mathrm{in}}_{2\,\omega} 
  \label{eq:12s}
 \end{align}
which correspond to Eq.~(3) of the main text. 

\section{Equations of motion: perturbative solution}
\label{sec:equat-moti-pert}
In the following we derive the solution of the equations of motion around
$\omega \simeq 0$, treating the probing tones $\mathcal{G}_{\Delta1}$ and
$\mathcal{G}_{\Delta2}$ perturbatively. The solution that we obtain for the
output field is then used to determine the output noise spectrum, from which, in
turn, it is possible to deduce the  mechanical
modes dynamics.  The solution of the equations of motion \eqref{eq:12s},
considering that
$\mathcal{G} \ll
\left|\mathcal{G}_{\Delta1}\right|=\left|\mathcal{G}_{\Delta2}\right|$, can be
obtained with the aid of perturbation theory. Defining
\begin{align*}
  \mathcal{G}_{\Delta1}= \lambda e^{\phi_1}\mathcal{G}_D \\
  \mathcal{G}_{\Delta2}= \lambda e^{\phi_2}\mathcal{G}_D
\end{align*}
we can formally write the solution for $a_\omega$,  $\beta_{1\,\omega}$, $\beta_{2\,\omega}$ as 
\begin{align}
 & a_\omega=a_\omega^{(0)} + \lambda a_\omega^{(1)} + \lambda^2 a_\omega^{(2)} + O(\lambda^3) 
   \nonumber \\ \nonumber \\
 & \beta_{\omega,1}=\beta_{\omega,1}^{(0)} + \lambda \beta_{1\,\omega}^{(1)} + \lambda^2 \beta_{1\,\omega}^{(2)} + O(\lambda^3)
   \nonumber \\ \nonumber \\
 & \beta_{2\,\omega}=\beta_{2\,\omega}^{(0)} + \lambda \beta_{2\,\omega}^{(1)} + \lambda^2 \beta_{2\,\omega}^{(2)} + O(\lambda^3).
  \label{eq:1x}
\end{align}
Substituting the perturbative expression given by Eq.~\eqref{eq:1x}, Eqs. \eqref{eq:12s} can be solved order-by-order in
$\lambda$,
\begin{widetext}
\begin{align}
    & {\left(\chi^{\mathrm{c}}_{\omega+\delta}\right)}^{-1}  a^{(n)}_\omega =
 - i \mathcal{G} \left[\beta^{(n)}_{1\,\omega} +\beta^{(n)}_{2\,\omega}\right] 
 -i  \lambda \mathcal{G}_D 
   \left\{
      e^{i\phi1}  
           \left[ \beta^{(n-1)}_{1\,\omega-\delta} +  \beta^{(n-1)}_{2\,\omega-\delta}\right]+ 
      e^{i \phi2}  
           \left[ \beta^{(n-1)}_{1\,\omega+\delta} +\beta^{(n-1)}_{2\,\omega+\delta} \right] 
   \right\}
+ \sqrt{\kappa_\mathrm{e}} a^{\mathrm{in}}_{\mathrm{E}\,\omega}  + \sqrt{\kappa_\mathrm{i}} a^{\mathrm{in}}_{\mathrm{I}\,\omega} \nonumber \\
\nonumber \\
&{\left(\chi^{\mathrm{m}}_{\omega-\delta}\right)}^{-1} \beta^{(n)}_{1\omega} = 
 - i \mathcal{G}^* a^{(n)}_\omega +
                         i \lambda   \mathcal{G}_D\left[  e^{-i\phi1} a^{(n-1)}_{\omega-\delta} + 
                                     e^{-i\phi2}  a^{(n-1)}_{\omega+\delta}\right] + \sqrt{\gamma} \beta^{\mathrm{in}}_{1\,\omega}\nonumber \\
\nonumber \\
&{\left(\chi^{\mathrm{m}}_{\omega+\delta}\right)}^{-1}  \beta^{(n)}_{2\omega} = 
 - i \mathcal{G}^*  a^{(n)}_\omega +
                        i \lambda \mathcal{G}_D\left[ e^{-i\phi1}   a^{(n-1)}_{\omega-\delta}+
                            e^{-i\phi2}   a^{(n-1)}_{\omega+\delta}\right] + \sqrt{\gamma} \beta^{\mathrm{in}}_{2\,\omega}.
  \label{eq:2x}
\end{align}
\end{widetext}
The $0$-th order term of  Eq.~\eqref{eq:2x} is represented by the equations governing the 2-pump
driving scheme, in the absence of detection tones which can be readily solved to give 
\begin{align}
  &a_\omega^{(0)} = \chi^\mathrm{c}_{\omega+\delta} \left[-i \mathcal{G} \left(\beta_{\omega,1}^{(0)}+\beta_{\omega,2}^{(0)} \right)+
                                                                              \sqrt{\kappa_\mathrm{e}} a^{\mathrm{in}}_{\mathrm{E}\,\omega}+ 
                                                                               \sqrt{\kappa_\mathrm{i}} a^{\mathrm{in}}_{\mathrm{I}\,\omega}
                                                                      \right] \nonumber \\\nonumber \\
  &\beta^{(0)}_{1\,\omega}=\chi_\omega^{\mathrm{e}1} \sqrt{\gamma}\beta^{\mathrm{in}}_{1\,\omega} - i \mathcal{G}
  \chi_\omega^{\mathrm{x}1}  \left( \sqrt{\kappa_\mathrm{e}} a^{\mathrm{in}}_{\mathrm{E}\,\omega} + 
                                               \sqrt{\kappa_\mathrm{i}} a^{\mathrm{in}}_{\mathrm{I}\,\omega}\right) \nonumber \\\nonumber \\
  &\beta^{(0)}_{2\,\omega}=\chi_\omega^{\mathrm{e}2} \sqrt{\gamma} \beta^{\mathrm{in}}_{2\,\omega} - i \mathcal{G}
  \chi_\omega^{\mathrm{x}2} \left( \sqrt{\kappa_\mathrm{e}} a^{\mathrm{in}}_{\mathrm{E}\,\omega} + 
                                               \sqrt{\kappa_\mathrm{i}} a^{\mathrm{in}}_{\mathrm{I}\,\omega}\right)
  \label{eq:3x}
\end{align}
where 
\begin{align}
  \chi^{\mathrm{e}1}_\omega =& \frac{ \chi^\mathrm{m}_{\omega-\delta}}
                                    {1+ \mathcal{G}^2 \chi^\mathrm{c}_{\omega+\delta} \chi^\mathrm{m}_{\omega-\delta}} 
\nonumber \\
 \chi^{\mathrm{e}2}_\omega =& \frac{ \chi^\mathrm{m}_{\omega+\delta}}
                                    {1+ \mathcal{G}^2 \chi^\mathrm{c}_{\omega+\delta} \chi^\mathrm{m}_{\omega+\delta}} 
\nonumber \\
  \chi^{x1}_\omega =& \frac{ \chi^\mathrm{c}_{\omega+\delta} \chi^\mathrm{m}_{\omega-\delta}}
                                    {1+ \mathcal{G}^2 \chi^\mathrm{c}_{\omega+\delta} \chi^\mathrm{m}_{\omega-\delta}}
\nonumber \\
  \chi^{\mathrm{x}2}_\omega =& \frac{ \chi^\mathrm{c}_{\omega+\delta} \chi^\mathrm{m}_{\omega+\delta}}
                                    {1+ \mathcal{G}^2 \chi^\mathrm{c}_{\omega+\delta} \chi^\mathrm{m}_{\omega+\delta}}
  \label{eq:21s}
\end{align}
The assumption $\gamma\ll \delta$ allows us to recognise that  $\beta^0_{1\,\omega}$ and $\beta^0_{2\,\omega}$ are
peaked around $\omega \simeq \delta$ and $\omega \simeq -\delta$ respectively, while $a_\omega^0$ exhibits a
double-peak structure for $\omega\simeq \pm \delta$. 

The solution given by \eqref{eq:3x}, allows us to write the first-order approximation of the system equations of motion
($n=1$ in Eq.~\eqref{eq:2x}). The considerations concerning the peak structure of the $0$-th order equations -
essentially because $\chi_\mathrm{m}\left(\omega\right) \sim \delta_\omega$ --, allow us to write the first order
approximation for the cavity field and the mechanics as
\begin{enumerate}
\item $\omega \simeq 0$
  \begin{align}
    &a_\omega^1 = -i \lambda \mathcal{G}_D\chi^\mathrm{c}_{\omega+\delta}  
      \left[e^{i \phi1} \beta_{\omega-\delta,2}^0 +e^{i \phi2} \beta_{\omega+\delta,1}^0   \right] \nonumber \\ \nonumber \\
    &\beta^1_{1\,\omega} = 0 \nonumber \\ \nonumber \\
    &\beta^1_{2\,\omega} = 0 \nonumber \\ \nonumber \\ 
    \label{eq:4x}
   \end{align}
The value of the first-order correction to the cavity field, can be expressed in terms of the input field
\begin{widetext}
   \begin{align}
     \label{eq:4dx}
      &a_\omega^{(1)} = -i \lambda \mathcal{G}_D\chi^\mathrm{c}_{\omega+\delta} \left[ e^{i \phi1}
  \left\{ \chi_{\omega-\delta}^{\mathrm{e}2} \sqrt{\gamma} \beta^{\mathrm{in}}_{\omega-\delta,2}
   - i \mathcal{G}
  \chi_{\omega-\delta}^{\mathrm{x}2}  \left( \sqrt{\kappa_i} a^{\mathrm{in}}_{\mathrm{I}\,\omega-\delta} + 
                                               \sqrt{\kappa_e} a^{\mathrm{in}}_{\mathrm{E}\,\omega-\delta}
                                       \right)
    \right\} \right.+\nonumber \\
 &\hphantom{a_\omega^{(1)} = -i \lambda \mathcal{G}_D\chi^\mathrm{c}_{\omega+\delta} \big[}
  \left.  e^{i \phi2}
  \left\{ \chi_{\omega+\delta}^{\mathrm{e}1} \sqrt{\gamma} \beta^{\mathrm{in}}_{\omega+\delta,1}
   - i \mathcal{G}
  \chi_{\omega+\delta}^{\mathrm{x}1}  \left( \sqrt{\kappa_i} a^{\mathrm{in}}_{\mathrm{I}\,\omega+\delta} + 
                                               \sqrt{\kappa_e} a^{\mathrm{in}}_{\mathrm{E}\,\omega+\delta}
                                       \right)
    \right\}
    \right].
   \end{align}
\end{widetext}
On the other hand, Eq.~\eqref{eq:4x} can be shown to encode the relevant
information about the mechanical quadratures
\begin{align}
  a_\omega^{(1)} =& -i \lambda \mathcal{G}_D\chi^\mathrm{c}_{\omega+\delta} \frac{e^{i \Phi}}{\sqrt{2}} 
\nonumber \\
                     & \left\{\left(u+v\right)
                      \left[
                              \cos \varphi \left( \bar{X}_{1\,\omega}+\bar{X}_{2\,\omega}\right) 
 \right. \right. \nonumber \\ &\left.
                    \hp{(u+v)} 
                      -\sin \varphi  \left( \bar{Y}_{1\,\omega} -\bar{Y}_{2\,\omega}\right)
                     \right]   \nonumber \\
                    & +\left(u-v\right)
                     \left[
                               \cos \varphi \left( \bar{Y}_{2\,\omega }+\bar{Y}_{2\,\omega}\right) 
\right. \nonumber \\ &\left. \left.
                   \hp{(u-v)} 
                             +\sin \varphi  \left( \bar{X}_{1\,\omega } -\bar{X}_{2\,\omega}\right)
                     \right]  
                       \right\} 
 \label{eq:4d2x}
\end{align}
where $\Phi=\phi_1+\phi_2$ and $\varphi=\phi_1-\phi_2$.

From Eq.~\eqref{eq:4dx} it is possible to notice that $a_\omega^1$ does not depend on the value of the input fields at
$\omega=0$, implying that the backaction term, to this order, will not give rise to any interference contribution with the
zeroth-order cavity field.    
\onecolumngrid
\item $\omega \simeq \delta$
  \begin{align}
   & a_\omega^{(1)} = - \lambda \mathcal{G}_D \frac{ \mathcal{G} \chi^\mathrm{c}_{\omega+\delta}\chi^\mathrm{m}_{\omega-\delta}}
                                    {1+\mathcal{G}^2 \chi^\mathrm{c}_{\omega+\delta}\chi^\mathrm{m}_{\omega-\delta}}
                            \left( e^{-i \phi_1} a_{\omega-\delta}^{(0)} + e^{-i \phi_2} a_{\omega+\delta}^{(0)} \right)
                                 \nonumber \\ \nonumber \\
    & \beta^{(1)}_{1\,\omega} = -i \lambda \mathcal{G}_D^* \chi^\mathrm{m}_{\omega-\delta}\left[1-\frac{\mathcal{G}^2
    \chi^\mathrm{c}_{\omega+\delta}\chi^\mathrm{m}_{\omega-\delta}}
        {1+\mathcal{G}^2 \chi^\mathrm{c}_{\omega+\delta}\chi^\mathrm{m}_{\omega-\delta}}\right]
             \left( e^{-i \phi_1} a_{\omega-\delta}^{(0)} + e^{-i \phi_2} a_{\omega+\delta}^{(0)} \right) \nonumber \\ \nonumber \\
    & \beta^{(1)}_{2\,\omega} = 0 \nonumber \\ \nonumber \\ 
    \label{eq5x}
   \end{align}
\item $\omega \simeq -\delta$
  \begin{align}
   & a_\omega^{(1)} = -  \lambda \mathcal{G}_D \frac{ \mathcal{G}\chi^\mathrm{c}_{\omega+\delta}\chi^\mathrm{m}_{\omega+\delta}}
                              {1+\mathcal{G}^2 \chi^\mathrm{c}_{\omega+\delta}\chi^\mathrm{m}_{\omega+\delta}}
                            \left( e^{-i \phi_1} a_{\omega-\delta}^{(0)} +e^{-i \phi_2} a_{\omega+\delta}^{(0)} \right)
                                 \nonumber \\ \nonumber \\
   &  \beta^{(1)}_{1\,\omega} = 0 \nonumber \\ \nonumber \\ 
   &  \beta^{(1)}_{2\,\omega} = -i \lambda \mathcal{G}_D^* \chi^\mathrm{m}_{\omega+\delta}\left[1-\frac{\mathcal{G}^2
    \chi^\mathrm{c}_{\omega+\delta}\chi^\mathrm{m}_{\omega+\delta} }
        {1+\mathcal{G}^2 \chi^\mathrm{c}_{\omega+\delta}\chi^\mathrm{m}_{\omega+\delta}}\right]
                \left( e^{-i \phi_1} a_{\omega-\delta}^{(0)} + e^{-i \phi_2} a_{\omega+\delta}^{(0)} \right)\nonumber \\ \nonumber \\
    \label{eq5dx}
   \end{align}
\twocolumngrid
\end{enumerate}
\onecolumngrid
The same strategy can be applied to determine the second-order approximation to the cavity field around $\omega \simeq
0$ ($n=2$ in Eq.~\eqref{eq:2x}), giving  
\begin{align}
  a_\omega^{(2)} = -i \lambda \mathcal{G}_D \chi^\mathrm{c}_{\omega+\delta}
                        \left( e^{i \phi_1} \beta_{2\,\omega-\delta}^{(1)}
                             +e^{i \phi_2} \beta_{1\,\omega+\delta}^{(1)}  \right)
  \label{eq:6x}
\end{align}
and, in terms of $0$-th order approximation,
\begin{align}
  a_\omega^{(2)} = - \lambda^2 \mathcal{G}_D^2 \chi^\mathrm{m}_{\omega}
                         &\left\{
                                \left[1-\frac{ \mathcal{G}^2 \chi^\mathrm{c}_{\omega}\chi^\mathrm{m}_{\omega}}
                                        {1+\mathcal{G}^2\chi^\mathrm{c}_{\omega}\chi^\mathrm{m}_{\omega}}\right]
                                \left( a_{\omega-2\delta}^{(0)} + e^{i(\phi_1-\phi_2)}a_\omega^{(0)}\right) \right.\nonumber \\
                           + & \hp{\Bigg[} \left.
                             \left[1-\frac{\mathcal{G}^2\chi^\mathrm{c}_{\omega+2\delta}\chi^\mathrm{m}_{\omega}}
                             {1+\mathcal{G}^2\chi^\mathrm{c}_{\omega+2\delta}\chi^\mathrm{m}_{\omega}}\right]
                                \left(  e^{i(\phi_2-\phi_1)} a_{\omega}^{(0)} +a_{\omega+2\delta}^{(0)}\right)
                         \right\}.
  \label{eq:7x}
\end{align}
As previously discussed, due to the structure of the mechanical response, the $0$-th order cavity response will be
peaked around $w \simeq \pm \delta$, implying that the terms appearing in Eq.~\eqref{eq:7x}, for $\omega \simeq 0$ are
almost solely determined by the input fields, allowing us to approximate 
\begin{align}
  a_\omega^{(2)} = - \lambda^2 \mathcal{G}_D^2 \chi^\mathrm{m}_{\omega}
                         & \left\{ \left[1-\frac{\mathcal{G}^2\chi^\mathrm{c}_{\omega}\chi^\mathrm{m}_{\omega}}
                                   {1+\mathcal{G}^2\chi^\mathrm{c}_{\omega}\chi^\mathrm{m}_{\omega}}\right]
                                \left[ \chi^\mathrm{c}_{\omega-\delta}\left( \sqrt{\kappa_\mathrm{e}} a^{\mathrm{in}}_{\mathrm{E}\,\omega-2\delta} + 
                                                                                            \sqrt{\kappa_\mathrm{i}}
                           a^{\mathrm{in}}_{\mathrm{I}\,\omega-2\delta}\right) + \right.\right. \nonumber \\
                   &   \hp{\Big\{ \big[ \left[1-\frac{\mathcal{G}^2\chi^\mathrm{c}_{\omega}\chi^\mathrm{m}_{\omega}}
                                  {1+\mathcal{G}^2\chi^\mathrm{c}_{\omega}\chi^\mathrm{m}_{\omega}}\right]}
                       \left.    \chi^\mathrm{c}_{\omega+\delta} e^{i(\phi_1-\phi_2)} \left( \sqrt{\kappa_\mathrm{e}} a^{\mathrm{in}}_{\mathrm{E}\,\omega} + 
                                                                                \sqrt{\kappa_\mathrm{i}}     a^{\mathrm{in}}_{\mathrm{I}\,\omega}\right)\right] +\nonumber \\
                             & \hp{\Bigg[} \left[1-\frac{\mathcal{G}^2\chi^\mathrm{c}_{\omega+2\delta}\chi^\mathrm{m}_{\omega}}
                                                      {1+\mathcal{G}^2 \chi^\mathrm{c}_{\omega+2\delta}\chi^\mathrm{m}_{\omega}}\right]
                                 \left[\chi^\mathrm{c}_{\omega+\delta} e^{i(\phi_2-\phi_1)} \left( \sqrt{\kappa_\mathrm{e}} a^{\mathrm{in}}_{\mathrm{E}\,\omega} + 
                                                                                               \sqrt{\kappa_\mathrm{i}}
                               a^{\mathrm{in}}_{\mathrm{I}\,\omega}\right)+ \right. \nonumber \\
                        &\hp{\Bigg[ \left[1-\frac{\mathcal{G}^2\chi^\mathrm{c}_{\omega+2\delta}\chi^\mathrm{m}_{\omega}}
                                                      {1+\mathcal{G}^2 \chi^\mathrm{c}_{\omega+2\delta}\chi^\mathrm{m}_{\omega}}\right]\big[}          
      \left. \left.   \chi^\mathrm{c}_{\omega+3 \delta} \left( \sqrt{\kappa_\mathrm{e}} a^{\mathrm{in}}_{\mathrm{E}\,\omega+2\delta} + 
                                                                                \sqrt{\kappa_\mathrm{i}}
                          a^{\mathrm{in}}_{\mathrm{I}\,\omega+2\delta}\right)\right]  \right\}.
  \label{eq:8x}
\end{align}
\twocolumngrid

In Eq.~\eqref{eq:8x}, contrary to first-order case, the term $a^{\mathrm{in}}_{\mathrm{E}\,\omega}$ can give rise to a
non-vanishing interference contribution.  Focusing only on the terms proportional to $a^{\mathrm{in}}_{\mathrm{I}\,\omega}$, and
assuming that $\delta \ll \kappa$,Eq.~\eqref{eq:8x} can be further approximated to give
\begin{align}
  a_\omega^{(2)} \simeq& - 2 \lambda^2 \mathcal{G}_D^2  \chi^\mathrm{c}_{\omega+\delta} \chi^\mathrm{m}_{\omega}
                    \left[1-\frac{\mathcal{G}^2\chi^\mathrm{c}_{\omega+\delta}\chi^\mathrm{m}_{\omega}}
                                  {1+\mathcal{G}^2\chi^\mathrm{c}_{\omega+\delta}\chi^\mathrm{m}_{\omega}}\right]
  \nonumber \\
                        & \cos\left[\phi_2-\phi_1 \right]
                       \left(\sqrt{\kappa_\mathrm{e}} a^{\mathrm{in}}_{\mathrm{E}\,\omega} + 
                              \sqrt{\kappa_\mathrm{i}}  a^{\mathrm{in}}_{\mathrm{I}\,\omega}\right) .
  \label{eq:8dx}
\end{align}
From Eqs. \eqref{eq:3x}, \eqref{eq:4x},\eqref{eq:8x}, it is thus possible to evaluate the output field quadratures (up to second order in
  the perturbative expansion discussed above). Setting $\lambda=1$ we have 
\begin{align}
  \label{eq:2nn}
  X_\omega^{\mathrm{out\,\theta}} &= \left[a_\omega^{\mathrm{out\,(0)}} + 
                                                       \sqrt{\kappa_\mathrm{e}}\left(a_\omega^{(1)} +a_\omega^{(2)}\right)
  \right]e^{i \theta}+ \mathrm{h.c.}, \omega \to -\omega \nonumber \\
                                             & \doteq  X_{\omega }^{(0)\,\mathrm{out\,\theta}} 
      + \sqrt{\kappa_\mathrm{e}} \left( X_{\omega}^{(1)\,\theta} + X_{\omega}^{(2)\,\theta}\right)
\end{align}
with 
\begin{align}
  &a_\omega^{(0)\,\mathrm{out}} = \sqrt{\kappa_\mathrm{e}} a_\omega^0 - a_\omega^\mathrm{in} \nonumber \\
  &X_{\omega}^{(0)\,\mathrm{out\,\theta}}= \left(a_\omega^{(0)\,\mathrm{out}} e^{-i\theta} + 
                                                               a_\omega^{(0)\,\mathrm{out}\,\dagger} e^{i \theta}
    \right)/\sqrt{2}  \nonumber \\
  &X_{\omega}^{(1)\,\theta}=\left(a_\omega^{(1)} e^{-i\theta} + 
                                                               a_\omega^{(1)\,\dagger} e^{i \theta}\right)/\sqrt{2}\nonumber \\
  &X_{\omega}^{(2)\,\theta}=\left(a_\omega^{(2)} e^{-i\theta} + 
                                                               a_\omega^{(2)\,\dagger} e^{i \theta}\right)/\sqrt{2}\nonumber \\
  \label{eq:9x}  
\end{align}
and analogously for the higher-order mechanical quadrature operators $X_{\omega 1}^{\mathrm{\theta}}$ and $X_{\omega
  2}^{\mathrm{\theta}}$.

Considering the thermal input discussed in the main text, and 
defining the spectrum for the output field as
\begin{align}
  S_\omega^{\mathrm{out}\,\theta} = \frac{1}{2}\left[
                       \braket{X_{-\omega}^{\mathrm{out\,\theta}}X_\omega^{\mathrm{out\,\theta}}}+
                       \braket{X_{\omega}^{\mathrm{out\,\theta}}X_{-\omega}^{\mathrm{out\,\theta}}}\right], 
  \label{eq:1}
\end{align}
with analogous definitions for each perturbative order, 
the relations given by Eq.~\eqref{eq:9x} allow us to write, up to second order in the
detection tone amplitude and for $\omega\simeq 0$, the spectrum of the output noise as
\begin{align}
   S_\omega^{\mathrm{out}\,\theta}  &= 
   S_{\omega}^{(0)\,\mathrm{out\,\theta}} +
   \kappa_\mathrm{e} S_{\omega}^{\mathrm{(1)}\,\theta}
 \nonumber \\ 
  &+ \sqrt{\kappa_\mathrm{e}}\left(
   \braket{X_{-\omega}^{(0)\,\mathrm{out}\,\theta} X_{\omega}^{(2)\,\mathrm{\theta}}} +
   \braket{X_{-\omega}^{(2)\,\mathrm{\theta}}X_{\omega}^{(0)\,\mathrm{out}\,\theta}}  
   \right).
  \label{eq:10x}
\end{align}
With the definitions given by Eqs.(\ref{eq:3x},\ref{eq:4x},~\ref{eq:8x}), Eq.~\eqref{eq:10x} can be written as
\begin{align}
    S_\omega^{\mathrm{out}\,\theta}=& 
  \frac{1}{2}\left[\braket{a_{\omega}^{(0)\,\mathrm{out}\,\dagger} a_\omega^{(0)\,\mathrm{out}}}+
  \braket{a_{-\omega}^{(0)\,\mathrm{out}} a_{-\omega}^{(0)\,\mathrm{out}\,\dagger}} \right. \nonumber \\
&+\kappa_\mathrm{e}  \braket{X_{-\omega}^{\mathrm{(1)}\,\theta} X_\omega^{\mathrm{(1)}\,\theta}}\nonumber \\
  &+
   \frac{\sqrt{\kappa_\mathrm{e}}}{2}\left( 
    \braket{a_{\omega}^{(0)\,\mathrm{out}\,\dagger} a_{\omega}^{(2)}}+ 
    \braket{a_{-\omega}^{(2)} a_{-\omega}^{(0)\,\mathrm{out}\,\dagger}} +
\right. \nonumber \\ &\left. \left. \hp{\frac{\sqrt{\kappa_\mathrm{e}}}{2}\big(+ }
     \braket{a_{\omega}^{(2)\,\dagger} a_{\omega}^{(0)\,\mathrm{out}}}+ 
    \braket{a_{-\omega}^{(0)\,\mathrm{out}} a_{-\omega}^{(2)\,\dagger}} \right)+
\right.  \nonumber  \\ 
& \left.  \hp{\frac{\sqrt{\kappa_\mathrm{e}}}{2}\big(+ } \omega  \to -\omega\right]
  \label{eq:11x} 
\end{align}
For $\omega \simeq 0$, we have that the $0$-th order term corresponds to the pure cavity response 
\begin{align}
  B^{\mathrm{in}}&=\frac{1}{2}\left(\braket{a_{\omega}^{(0)\,\mathrm{out}\,\dagger} a_\omega^{(0)\,\mathrm{out}}}+ 
  \braket{a_{-\omega}^{(0)\,\mathrm{out}} a_{-\omega}^{(0)\,\mathrm{out}\,\dagger}} \right)
 \nonumber \\
  &\simeq 
  \left|\chi_{\delta}^\mathrm{c}-1\right|^2 \left(n_\mathrm{E}+\frac{1}{2}\right) +
  \kappa_\mathrm{e}\kappa_\mathrm{i} \left|\chi_{\delta}^\mathrm{c}\right|^2 \left(n_\mathrm{i}+\frac{1}{2}\right)
  \label{eq:12x}
\end{align}
while the terms appearing the third and fourth line  of Eq.~\eqref{eq:11x} can be written as
\begin{widetext}
\begin{align}
   C_\omega^{\mathrm{in}} \cos \left[ 2 \varphi \right]= \frac{\sqrt{\kappa_\mathrm{e}}}{2}&\left( 
   \braket{a_{\omega}^{(0)\,\mathrm{out}\,\dagger} a_{\omega}^{(2)}}+ 
   \braket{a_{-\omega}^{(2)} a_{-\omega}^{(0)\,\mathrm{out}\,\dagger}} +
   \braket{a_{\omega}^{(2)\,\dagger} a_{\omega}^{(0)\,\mathrm{out}}}+ 
   \braket{a_{-\omega}^{(0)\,\mathrm{out}} a_{-\omega}^{(2)\,\dagger}} \right)= \nonumber \\
     & \hp{+ 
   \braket{a_{-\omega}^{(2)} a_{-\omega}^{(0)\,\mathrm{out}\,\dagger}}}
    \cos 2 \varphi \left\{ 
    \operatorname{Re}\left[ \left(\chi_{-\omega+\delta}^\mathrm{c}-1\right) A_{-\omega}^{(2)}\right]
     \kappa_\mathrm{e} \left( n_\mathrm{e} +1\right)  +
  \operatorname{Re}\left[ \left(\chi_{\omega+\delta}^\mathrm{c}-1\right) A_\omega^{(2)}\right]
     \kappa_\mathrm{e} n_\mathrm{e}
                                       \right. \\
 & \hp{\cos 2 \varphi + 
   \braket{a_{-\omega}^{(2)} a_{-\omega}^{(0)\,\mathrm{out}\,\dagger}}}
\left.  
  + \operatorname{Re}\left[\chi_{-\omega+\delta}^\mathrm{c}A_{-\omega}^{(2)}\right]
     \sqrt{\kappa_\mathrm{e} \kappa_\mathrm{i}} \left( n_\mathrm{i} +1\right) +   
   \operatorname{Re}\left[\chi_{\omega+\delta}^\mathrm{c} A_\omega^{(2)}\right]
     \sqrt{\kappa_\mathrm{e} \kappa_\mathrm{i}} n_\mathrm{i}  
  \right\}
  \label{eq:14x}
\end{align}
where we have defined 
\begin{align}
  A_\omega^{(2)} = - 2 \mathcal{G}_D^2  \chi^\mathrm{c}_{\omega+\delta} \chi^\mathrm{m}_{\omega}
                    \left[1-\frac{\mathcal{G}^2\chi^\mathrm{c}_{\omega+\delta}\chi^\mathrm{m}_{\omega}}
                                  {1+\mathcal{G}^2\chi^\mathrm{c}_{\omega+\delta}\chi^\mathrm{m}_{\omega}}\right].
  \label{eq:13x}
\end{align}
The contribution to the output field noise spectrum given by
$\kappa_\mathrm{e} \braket{X_{-\omega}^{(1)\,\mathrm{\theta}}X_{\omega }^{(1)\,\mathrm{\theta}}}$ can be shown to encode
the relevant information about the mechanical quadratures. Again for $\omega \simeq 0$, from \eqref{eq:4d2x} we have that  
\begin{align}
  S_\omega^{\mathrm{(1)}\,\theta} \simeq
    \mathcal{G}^2_D \left|\chi_\mathrm{c}(\delta)\right|^2 
     &\left\{
            \left(u+v\right)^2 
          \left[
               \left(\cos \theta \cos \varphi \right)^2  \bar{S}_\omega^{\Sigma\,0} +
               \left(\cos \theta \sin \varphi \right)^2  \bar{S}_\omega^{\Delta\,\pi/2}
          \right] \right. \nonumber \\ 
        &
         \left.  + \left(u-v\right)^2  
          \left[
                 \left(\sin \theta \cos \varphi \right)^2  \bar{S}_\omega^{\Sigma\,\pi/2} +
               \left(\sin \theta \sin \varphi \right)^2    \bar{S}_\omega^{\Delta\,0}
          \right]
    \right\}
  \label{eq:14x2}
\end{align}
\end{widetext}
where 
\begin{align}
  &\bar{S}_\omega^{\Sigma\, 0}= \frac{1}{2}\left[
                       \braket{\bar{X}_{-\omega}^{\Sigma}\bar{X}_\omega^{\Sigma}}+
                       \braket{\bar{X}_{\omega}^{\Sigma} \bar{X}_{-\omega}^{\Sigma}}\right] \nonumber \\
 &\bar{S}_\omega^{\Sigma\,\pi/2}=\frac{1}{2}\left[\braket{\bar{Y}_{-\omega}^{\Sigma}\bar{Y}_\omega^{\Sigma}}+
                       \braket{\bar{Y}_{\omega}^{\Sigma}\bar{Y}_{-\omega}^{\Sigma}}\right] \nonumber \\
&\bar{S}_\omega^{\Delta\, 0}= \frac{1}{2}\left[
                       \braket{\bar{X}_{-\omega}^{\Delta}\bar{X}_\omega^{\Delta}}+
                       \braket{\bar{X}_{\omega}^{\Delta} \bar{X}_{-\omega}^{\Delta}}\right] \nonumber \\
  &\bar{S}_\omega^{\Delta\,\pi/2}=\frac{1}{2}\left[\braket{\bar{Y}_{-\omega}^{\Delta}\bar{Y}_\omega^{\Delta}}+
                       \braket{\bar{Y}_{\omega}^{\Delta}\bar{Y}_{-\omega}^{\Delta}}\right]
 \label{eq:3}
\end{align}

\section{Duan quantity}
\label{sec:duan-quantity-from}

From the QLEs equations for the mechanical Bogoliubov modes, we evaluate here
$\braket{\Delta \bar{X}_\Sigma^2} + \braket{\Delta \bar{Y}_\Delta^2} $, which,
as we will discuss in the next section, can be shown to correspond to the Duan
quantity. From Eqs. \eqref{eq:12s} (Eqs. (3) of the main
text), 
in the appropriate frame for each mode, we can write the I/O relations for
$\beta_1$ and $\beta_2$ as
\begin{align}
  \beta_{\omega,1} =& \chi^{e1}_{\omega-\delta} \sqrt{\gamma} \beta_{1}^{\mathrm{in}} -i \mathcal{G} \chi^{x1}_{\omega-\delta} \left(
\sqrt{\kappa_{\mathrm{i}}} a_{I}^{\mathrm{in}}+\sqrt{\kappa_{\mathrm{e}}}a_{E}^{\mathrm{in}}\right) \nonumber \\
  \beta_{\omega,2} =& \chi^{e2}_{\omega+\delta} \sqrt{\gamma} \beta_{2}^{\mathrm{in}} -i \mathcal{G} \chi^{x2}_{\omega+\delta} \left(
\sqrt{\kappa_{\mathrm{i}}} a_{I}^{\mathrm{in}}+\sqrt{\kappa_{\mathrm{e}}}a_{E}^{\mathrm{in}}\right) 
  \label{eq:20s}  
\end{align}

Through Eqs. (4) of the main text, Eqs. \eqref{eq:20s} can be expressed in terms of mechanical quadrature
operators
\begin{align}
    & \bar{X}_\omega^{\Sigma\,\Delta}=\left. \bar{X}_\omega^{\theta,\Sigma\,\Delta}\right|_{\theta=0}\nonumber \\
    & \bar{Y}_\omega^{\Sigma\,\Delta}=\left. \bar{X}_\omega^{\theta,\Sigma\,\Delta}\right|_{\theta=\pi/2}
  \label{eq:5}
\end{align}
and input operators
$b_{1\,\mathrm{in}}^{(\dagger)}$, $b_{2\,\mathrm{in}}^{(\dagger)}$, $a_{\mathrm{in}}^{(\dagger)}$ as 
\begin{align}
 \bar{X}_\omega^{\theta,\Sigma\,\Delta} =  \frac{1}{\sqrt{2}} &\left(A^{\Sigma\,\Delta}_{1\, \omega} b_{1\,\mathrm{in}} + A^{\Sigma\,\Delta}_{2\,\omega}
  b_{2\,\mathrm{in}} +   C^{\Sigma,\Delta}_\omega a_{\mathrm{in}} +
\right. \nonumber \\ 
& \left. \mathrm{h.c.}, \, \omega \to -\omega \right)
  \label{eq:22s}
\end{align}
with 
\begin{align}
  A^{\Sigma\,\Delta}_{1\, \omega}&= \sqrt{\gamma}
                                              \left[ 
                                                 \eta^\theta_\pm    u  \chi^\mathrm{e1}_{\omega} \pm 
                                                 {\eta^\theta_\pm}^* v \chi^\mathrm{e2*}_{-\omega} 
                                              \right] \nonumber \\
  A^{\Sigma\,\Delta}_{2\, \omega}&=\pm\sqrt{\gamma}
                                              \left[ 
                                                 \eta^\theta_\pm    u  \chi^\mathrm{e2}_{\omega} \pm 
                                                 {\eta^\theta_\pm}^* v \chi^\mathrm{e1*}_{-\omega} 
                                              \right]\nonumber \\
  C^{\Sigma,\Delta}_\omega&=-i \sqrt{k}\mathrm{G} 
                                          \left[
                                            \eta^\theta_\pm \left(\chi^\mathrm{x1}_{\omega}\pm 
                                                                         \chi^\mathrm{x2}_{\omega}   
                                                         \right) 
                                         \right]
  \label{eq:23s}
\end{align}
where $\eta^\theta_\pm= u e^{-i \theta}\mp v e^{i \theta}$.
From Eqs. \eqref{eq:22s} and \eqref{eq:23s}, and assuming $\braket{b^\dagger_1 b_1}=\braket{b^\dagger_2
  b_2}=n_\mathrm{m}$,  we can evaluate the quadrature variances for the symmetric and antisymmetric modes
\begin{align}
 \braket{\Delta \bar{X}_{\Sigma,\Delta}^{\theta\,2}} \int \frac{d\omega}{2 \pi}
  S_\omega^{\Sigma,\Delta \,\theta}
  \label{eq:6}
\end{align}
 as  
\begin{widetext}
\begin{align}
  \braket{\Delta \bar{X}_{\Sigma,\Delta}^{\theta\,2}}_\mathrm{m} &=
      \gamma  
                   \left\{ \left|\eta^\theta_\pm \right|^2  \left(u^2+v^2 \right) 
                  \left( 
                         \int \frac{d \omega}{2 \pi}  \left| \chi^\mathrm{e1}_{\omega} \right|^2+
                         \int \frac{d \omega}{2 \pi}  \left| \chi^\mathrm{e2}_{\omega} \right|^2
                  \right)
                  \pm  
                  4 u v \operatorname{Re} 
                 \left[
                    {\eta^\theta_\pm}^2    \int \frac{d \omega}{2 \pi}  \chi^\mathrm{e1}_{\omega} \chi^{\mathrm{e2}}_{\omega}
                 \right]\right\} \left( n_\mathrm{m}+\frac{1}{2}\right)\nonumber \\
  \braket{\Delta \bar{X}_{\Sigma,\Delta}^{\theta\,2}}_\mathrm{o} &=
      \kappa \mathcal{G}^2 \left|\eta^\theta_\pm \right|^2
                            \left\{
                            \left|\eta^\theta_\pm \right|^2  \int \frac{d \omega}{2 \pi}  \left| \chi^{\mathrm{x1}}_{\omega} \right|^2+
                          \int \frac{d \omega}{2 \pi}  \left| \chi^{\mathrm{x2}}_{\omega} \right|^2 \pm 
                            2 \operatorname{Re} 
                  \left[
                      {\eta^\theta_\pm} ^2 \int \frac{d \omega}{2 \pi}  \chi^\mathrm{x1}_{\omega} \chi^{\mathrm{x2}}_{\omega}
                 \right]  \right\}  \left(n_\mathrm{c}+\frac{1}{2}\right)    
   \label{eq:24s}
\end{align}
where we have separated the contributions that can be ascribed to the mechanical $\braket{\cdot}_\mathrm{m}$ and the
optical $\braket{\cdot}_\mathrm{c}$ thermal bath. The response integrals can be evaluated analytically, giving (in the
limit $\gamma \ll \delta$) 
\begin{align}
    \braket{\Delta \bar{X}_{\Sigma}^2}_{\mathrm{m}} = \braket{\Delta {\bar{Y}_{\Sigma}}^2}_\mathrm{m} 
           =\frac{\gamma \left(u-v\right)^2 \left(8 \delta ^2+4 G^2+\kappa ^2\right)
                                                             }{2 G^2\kappa} \left[\left(u^2+v^2\right) 
                            +  \frac{\kappa ^2 }{ 2 \left( \delta ^2  +\kappa^2/4 \right)} uv \right]
                   ( n_\mathrm{m}+\frac{1}{2}) 
  \label{eq:25s}
\end{align}
\end{widetext}
and
\begin{align}
   \braket{\Delta \bar{X}_{\Sigma}^2}_\mathrm{o}= \braket{\Delta {\bar{Y}_{\Sigma}}^2}_\mathrm{o}=\left(u-v\right)^2  \left(n_\mathrm{c}+\frac{1}{2}\right)  
  \label{eq:26s}
\end{align}
which, for $\delta \ll \kappa$, allows us to recover the result given in Eq.~(14) of the main 
text. 

\section{Frequency-shifted quadratures and Duan bound}
\label{sec:freq-shift-quadr}

In order to confirm the presence entanglement between the mechanical resonators, we have
to show that the variances of symmetric and antisymmetric quadratures satisfy
the Duan bound, which can be expressed in the following form
\begin{align}
   \braket{\Delta {X^\theta_\Sigma}^2} +\braket{\Delta {X^{\theta+\pi/2}_\Delta}^2} \leq 1.
   \label{eq:13s}
\end{align}
where  
\begin{align}
  &\braket{\Delta {X^\theta_\Sigma}^2} =\int \frac{d\omega}{2 \pi}
  S_\omega^{\Sigma\,\theta} \nonumber \\
  &\braket{\Delta {X^\theta_\Delta}^2} =  \int \frac{d\omega}{2 \pi}
  S_\omega^{\Sigma\,\theta+\pi/2} 
  \label{eq:14s}
\end{align}
with 
\begin{align}
  X^\theta_{\Sigma,\Delta}= X^\theta_{\omega,1} \pm  X^\theta_{\omega,2}
  \label{eq:15s}  
\end{align}
where the upper (lower) sign corresponds to $X_\omega^{\theta, \Sigma}$
($X_\omega^{\theta, \Delta}$) and 
\begin{align}
  X^\theta_{\omega,1} =\left(b^\dagger_{-\omega,1}e^{i \theta}
                               +b_{\omega,1}e^{-i\theta}\right)/\sqrt{2} \nonumber \\
  X^\theta_{\omega,2}=\left(b^\dagger_{-\omega,2}e^{i \theta}
                               +b_{\omega,2}e^{-i\theta}\right)\sqrt{2}.
  \label{eq:16s}
\end{align}
While the quadratures given in eq. \eqref{eq:15s} cannot be directly related to the mechanical contribution to the
output spectrum, given by $\bar{X}^\theta_{\omega,1}$ and
$\bar{X}^\theta_{\omega,2}$ (see Eq.~(9) of the main text), it is possible to show that the following identity holds 
\begin{align}
  \int \frac{d\omega}{2 \pi} \left[\bar{S}_\omega^{\Sigma\, \theta} +\bar{S}_\omega^{\Delta\, \theta+\pi/2}  \right] = 
  \int \frac{d\omega}{2 \pi} \left[S_\omega^{\Sigma\,  \theta} +S_\omega^{\Delta \, \theta+\pi/2}\right].
  \label{eq:17s}  
\end{align}
In order to prove Eq.~\eqref{eq:17s}, we consider, without loss of generality the case $\theta=0$.
The argument of the integral in Eq.~\eqref{eq:17s}
can be written in terms of the definitions given in eq. \eqref{eq:6s} as 
 \begin{align}
 & \braket{\left(\bar{X}_{-\omega,1}+\bar{X}_{-\omega,2}\right)\left(\bar{X}_{\omega,1}+\bar{X}_{\omega,2}\right)} 
 \nonumber \\
&+ 
   \braket{\left(\bar{Y}_{-\omega,1}-\bar{Y}_{-\omega,2}\right)\left(\bar{Y}_{\omega,1}-\bar{Y}_{\omega,2}\right)}  = 
   \nonumber \\
   &\bigg(
    \braket{b^\dagger_{\omega+\delta,1}b^\dagger_{\omega+\delta,1}} +
   \braket{b_{-\omega+\delta,1}b^\dagger_{-\omega+\delta,1}} +  
  \braket{b^\dagger_{\omega+\delta,1}b^\dagger_{-\omega-\delta,2}} +  \nonumber \\ 
   & \braket{b_{-\omega+\delta,1}b_{-\omega-\delta,1}}+  
    \braket{b^\dagger_{\omega-\delta,2}b^\dagger_{-\omega+\delta,1}} +
   \braket{b_{-\omega-\delta,2}b_{\omega+\delta,2}}
   \bigg).  
 \label{eq:19s}
 \end{align}
 Since the integration has to be performed over the whole frequency domain, upon integration, the frequency in each term
 can be shifted by the appropriate amount (either $\omega \to \omega + \delta$ or $\omega \to \omega - \delta$),
 reproducing the result that would be obtained directly evaluating the integral appearing on the rhs of
 Eq.~\eqref{eq:17s}. It is thus clear that the evaluation of the output field spectrum given in Eq.~(9) of the main
 text 
 allows us to determine $S_\omega^{\Sigma\,\theta}$, $S_\omega^{\Sigma \theta}$, and consequently, the Duan
 quantity, therefore representing a measure of the degree of entanglement between the mechanical resonators.



\end{document}